\documentclass[aps,preprint,showpacs]{revtex4}
\usepackage{graphicx}
\begin{document}
\title{Unveiling the origin of shape coexistence in  lead isotopes}                
\author{J.L. Egido}
\author{L.M. Robledo}
\affiliation{Departamento de F\'{\i}sica
Te\'orica, Universidad Aut\'onoma de Madrid, E-28049 Madrid, Spain} 
\author{R.R. Rodr\'{\i}guez-Guzm\'an}
\affiliation{ Institut f\"{u}r Theoretische Physik der Universit\"{a}t T\"{u}bingen,
 D-72076 T\"{u}bingen, Germany. }
\date{\today}

\begin{abstract}
The shape coexistence in the nuclei $^{182-192}$Pb is analyzed within
the Hartree-Fock-Bogoliubov  approach with the effective Gogny force.
A good agreement with the experimental energies is found for the coexisting 
spherical, oblate and prolate states.  Contrary to the established interpretation,
it is found that the low-lying prolate and oblate  $0^+$  states 
observed in this mass region are predominantly characterized by neutron correlations
whereas the protons behave as spectators rather than playing an active role.
\end{abstract}

%

%

%
%
\maketitle

The experimental information about the phenomenon of shape coexistence in atomic 
nuclei has been rather scarce until recent years. Nowadays this topic
is one of the most  active research fields in Nuclear Physics
and rather widespread in the Nuclide Chart. Nowhere, however,
coexistence appears as clear and impressive as in the neutron deficient lead
isotopes (\( N\sim 104 \)) \cite{AHD.00}. There the ground state and the first
 two excited states are
$0^+$ states, all the three are within 1 MeV and each of them corresponds
to a different shape: the ground state is spherical, one excited state is
prolate and the other one oblate. The fact that this situation takes place in
several lead isotopes makes this region more attractive to theoretically
understand the phenomenon of coexistence.   Since many 
years there has been considerable interest in this region and quite a few 
theoretical studies have tried to disentangle the main features of this
coexistence. The older works were done in the frame of the Strutinsky method 
\cite{MPF.77,BN.89,HDR.89,NAZ.93} as well as in terms of a schematic
SU(3) shell model calculation with degenerate major shells \cite{HJM.87}.
Recent calculations beyond mean field \cite{CER.01,DBB.03} have studied
the stability of the mean field predictions. In particular, in Ref.~\cite{CER.01}
configuration mixing (shape fluctuations) calculations with the Gogny force
in the frame of the Generator Coordinate Method as well as many-body calculations
with a separable force have been performed.  In Ref.\cite{DBB.03}, angular momentum
conserving shape fluctuations were considered with the Skyrme interaction. 
 All three calculations confirmed  the persistence of the low-lying prolate and 
oblate  configurations  predicted in the mean field calculations. Unfortunately in
these calculations with effective interactions, not much interest was devoted to 
understand the physics underlying the phenomenon of coexistence. In this Letter we adopt
the opposite point of view: taking for granted that beyond mean field correlations
do not destroy the simpler mean field predictions, we will perform a detailed
study with the Hartree-Fock-Bogoliubov (HFB) approach and effective forces with predictive power
to pin down the subtilities of the shape coexistence.

   The ``traditional" interpretation of the low-lying prolate and oblate minima 
in the closed shell lead isotopes is based in the 2p-2h proton configurations
coming down in energy due to the pairing correlations and the proton-neutron (p-n)
interaction, see \cite{HJM.87,HDR.89}. This interpretation is still common sense today \cite{AHD.00}. 
In this Letter we will show that a different, much more natural,
interpretation of these states emerges in a full selfconsistent calculation with an 
effective interaction. The calculations have been performed within the HFB
approach using  the finite range density dependent Gogny force (D1S parametrization)\cite{BGG.84}.
Energies, $E(q)$, and wave functions (w.f.), $\varphi (q)$,  with deformation $q$, have 
been calculated by the constrained
HFB method,  by constraining the expectation value of the mass  quadrupole moment, i.e.,  
\( \left\langle \varphi (q)\right| z^{2}-1/2(x^{2}+y^{2})\left| \varphi (q)\right\rangle = q \). 
The calculations have been performed in an axially symmetric Harmonic Oscillator (HO) basis containing 
thirteen major shells, see also ref.~\cite{CER.01}.

\begin{figure}[h!]

{\centering {\includegraphics[angle=0,width=\columnwidth]{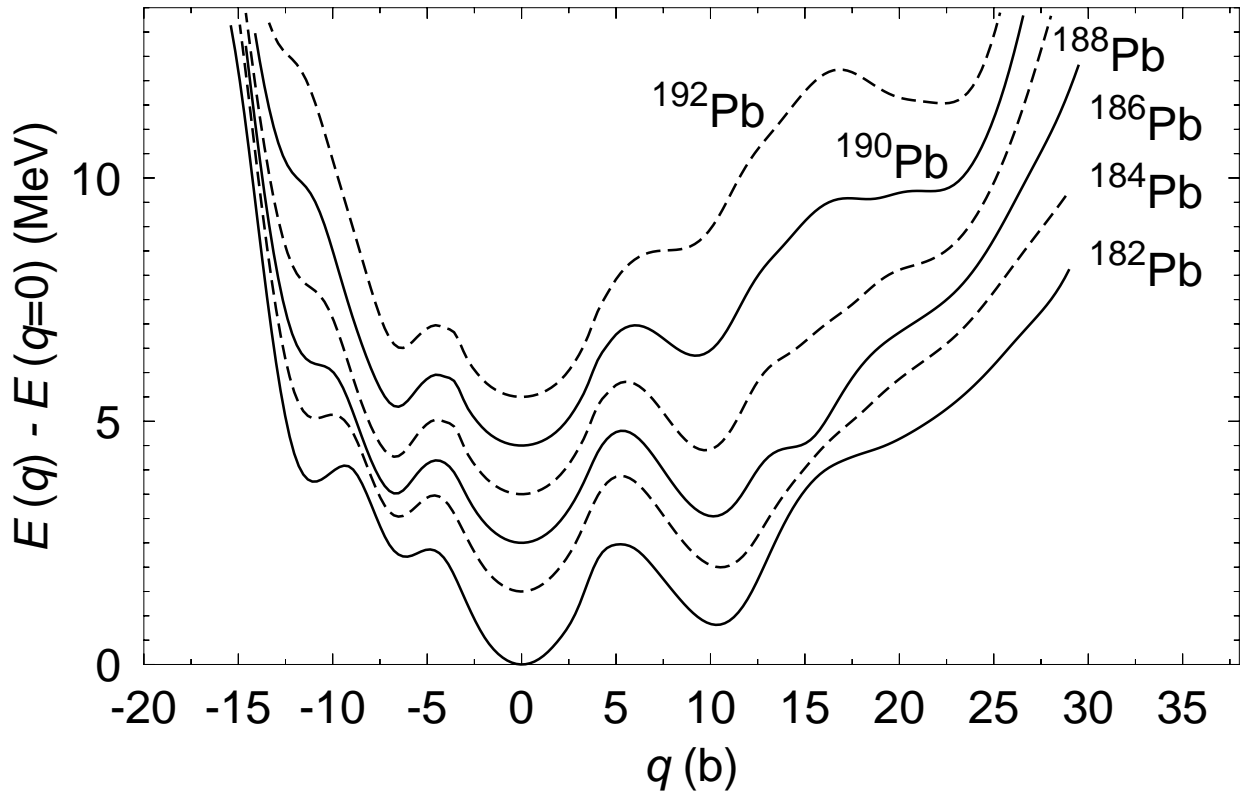}} \par}

\caption{Potential energies for the Pb isotopes as functions of $q$. 
The curves for  \protect\(^{184-192}\)Pb have been shifted
in  1.5, 2.5, 3.5, 4.5 and 5.5 MeV respectively.}\label{Fig:MFPES}
\end{figure}

\begin{figure}

{\centering {\includegraphics[angle=0,width=\columnwidth]{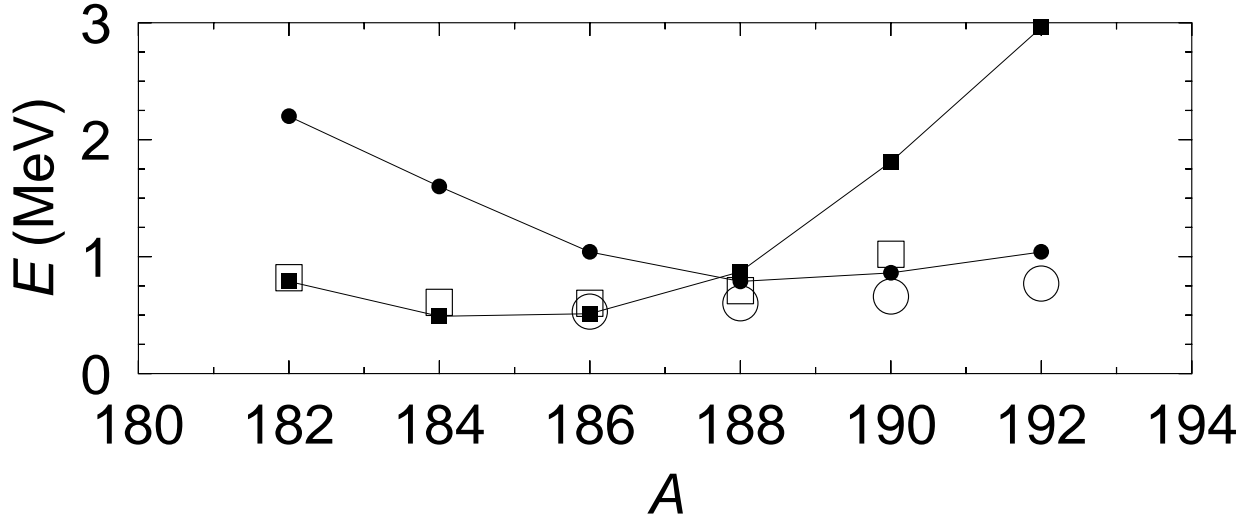}} \par}

\caption{Excitation energies of the oblate minima, filled (open) circles for theory (experiment),
and the prolate ones, filled (open) box for theory (experiment) for the nuclei
\protect\(^{182-192}\)Pb.}\label{Fig:MF_MIN}
\end{figure}

In Fig.~\ref{Fig:MFPES}, the  HFB energy as functions of  $q$
is displayed for the isotopes $^{182-192}$Pb. All these nuclei have 
spherical ground states,   giving us a clear 
indication of the  very strong signature of the Z=82 proton shell closure. 
Additional  low-lying minima around $10 $b on the prolate
and around $ -7 $b on the oblate side  appear indicating a pronounced shape coexistence.
The first (second) excited states in \(^{182-184-186}\)Pb are of prolate (oblate) shape with 
deformation parameters \({\beta}_{2}=0.29, 0.32\) and \(0.29\) (\(-0.18, -0.18\) and \(-0.20\))
respectively. 
For \(^{188-190-192}\)Pb, the relative positions of the deformed minima are 
exchanged. The first (second ) excited states  are of oblate (prolate)
shape with \({\beta}_{2}=-0.20, -0.20\) and \(-0.17\) (\(0.28, 0.25\) and \(0.22\)) 
 respectively. It is interesting to notice that the oblate minima become more pronounced
with increasing neutron number while the prolate ones display the opposite trend, in particular,
the prolate minimum in \(^{192}\)Pb is rather shallow. On the other hand, the changes taking 
place on the prolate side of the potential are stronger than on the oblate one.
The excitation energies of the lowest-lying oblate and prolate minima in 
\(^{182-192}\)Pb are compared in Fig. \ref{Fig:MF_MIN}
with the available experimental data  
\cite{JULI,HMG.93,AHD.00,APL.98,DUP.87,DCD.85,PB2,PB3,PB4,PB5}.
The crossing of the theoretical prolate and oblate states takes place at $A=188$, two units higher  
than suggested by the experiment (A=186)\cite{JULI}.  It is interesting to notice that the
agreement with the experiment is better for the first excited states than for the second ones
independently of their  oblate or prolate shape. A similar tendency  has been obtained 
in \cite{DBB.03}. 
The results obtained for the excitation energies, without further adjustment of the force, 
indicate that
though the mean field approximation does not reproduce quantitatively
the experimental results it provides a good qualitative description
of the three minima. We will now try to understand the physics behind them.

\begin{figure}

{\centering {\includegraphics[angle=-90,width=\columnwidth]{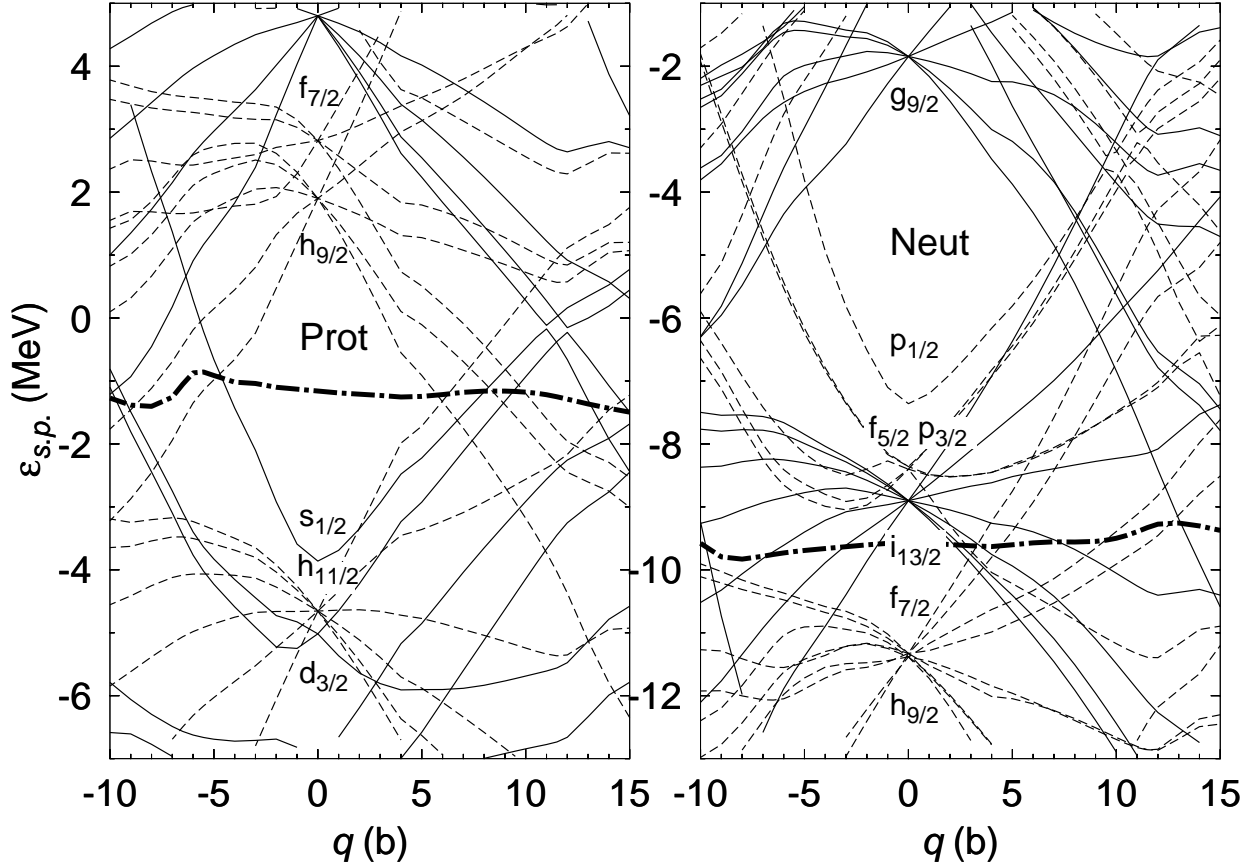}} \par}

\caption{Single particle energies of the nucleus \protect\(^{186}\)Pb. Dashed (full) lines 
correspond to negative  (positive) parity states. The thick lines represent the 
chemical potentials.}
\label{Fig:SPE}
\end{figure}
 
\begin{figure}

{\centering {\includegraphics[angle=0,width=\columnwidth]{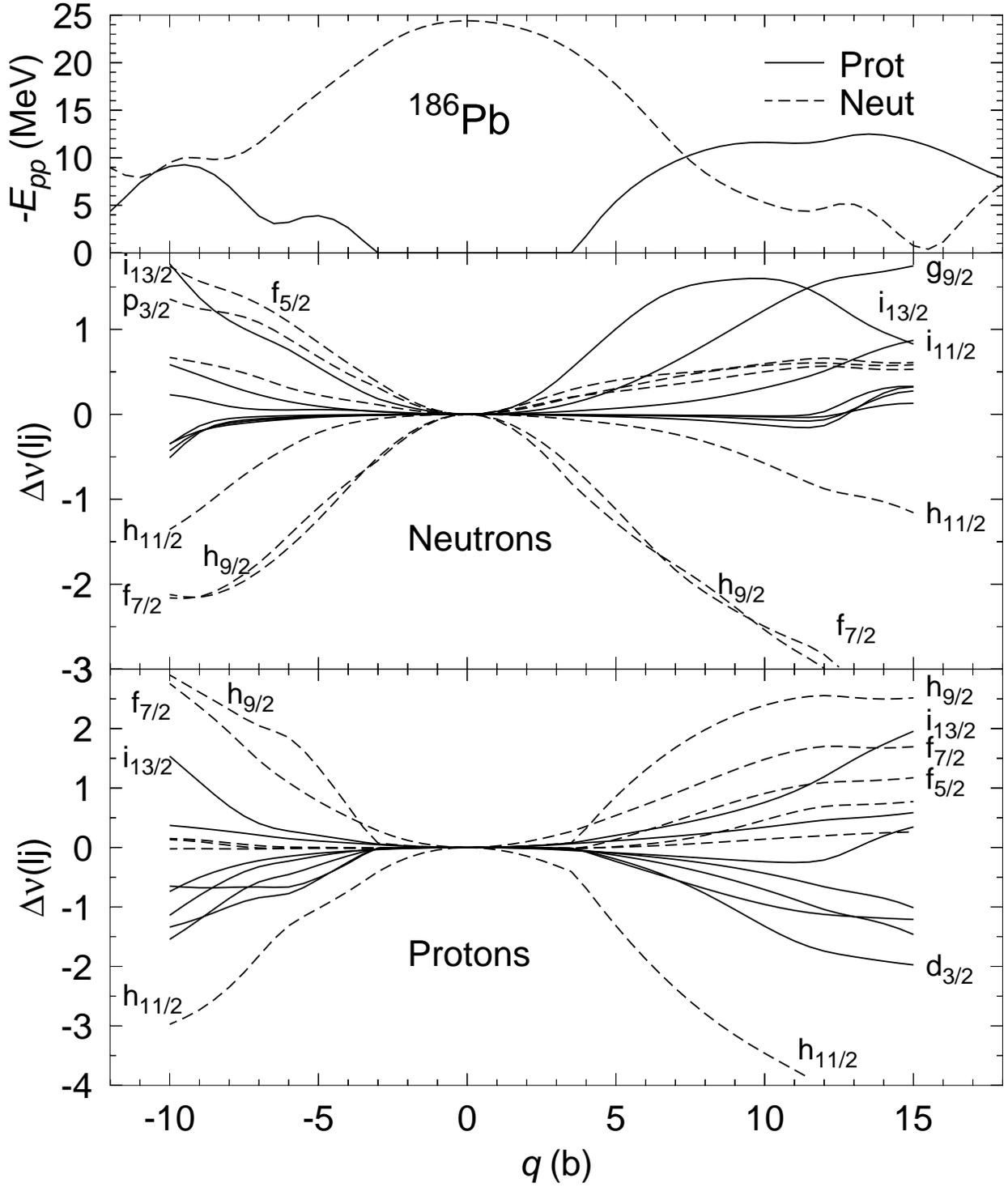}} \par}

\caption{ Lower panels: Relative single particle occupations of the levels around the 
Fermi surface for \protect\(^{186}\)Pb as a function of $q$. Top panel: Pairing
energies as a function of $q$.}
\label{Fig:occ}
\end{figure}

An interesting issue in mean field calculations are the single particle energies 
 (s.p.e.) corresponding to the eigenvalues  of the one-body Hamiltonian $h$.
 In Fig.~\ref{Fig:SPE} the s.p.e. for protons and neutrons around the Fermi level
are plotted for the nucleus  \(^{186}\)Pb as an example.  We observe that the first level crossings
in the Fermi surface appear  around $|3| $b for the neutron system and for the proton system, 
due to the $ Z = 82 $ shell closure,  around $|6| $b.
According to the assignment of earlier works \cite{HJM.87,NAZ.93} the prolate minimum lying around
10 b is a proton  4p-4h state, while the oblate one around $-7$ b is a proton 2p-2h. This
labeling is somewhat misleading because the reference state of the p-h excitations 
is changing with deformation. In the framework
of Fig.~\ref{Fig:SPE} it is more appropriated to talk about level crossings, one level crossing for the oblate
state and two for the prolate one. The neutron systems are characterized by more crossings due to the
fact that the first level crossing  already takes place at $|3| $b. For a genuine determination -in the
spirit of the shell model- of the p-h
excitations at a given $q$ one should calculate the single particle occupancies 
\( \nu (lj,q)  =  \langle \varphi(q) | \sum_{n,m}a^{\dagger}_{nljm} a^{}_{nljm} |\varphi(q)
\rangle\)  with $a^{\dagger}_{nljm}, a^{}_{nljm}$ the particle creation and annihilation operators of 
the {\em spherical}  HFB solution. 
For each $(lj)$ the quantity $\Delta \nu (lj)= \nu (lj,q) - \nu (lj,q=0)$ represents the 
relative spherical total single particle 
occupancy  of the orbital $lj$ at a given deformation $q$  with respect to the corresponding one at 
zero deformation and  is plotted in Fig.~\ref{Fig:occ} again  for the nucleus \(^{186}\)Pb. 
The proton w.f. of the prolate (oblate) minimum  corresponds to a 7p-7h (4.5p-4.5h) with respect to the 
spherical state. The 7p are distributed, mainly, within the $h_{9/2}, f_{7/2}$ and  $f_{5/2}$ orbitals
while the 7h are made, mainly,  in the $h_{11/2}$ orbital. The 4.5p-4.5h of the oblate minimum are distributed
within the same orbitals (with the exception of the $f_{5/2}$).
The neutron w.f. of the prolate (oblate) minimum  corresponds to a 5.5p-5.5h (4.4p-4.4h) with respect to the 
spherical state. The 5.5p are distributed, mainly, within the $i_{13/2}$ and the $g_{9/2}$  orbitals
while the 5.5h are made in the $h_{9/2}, f_{7/2}$ orbitals. The 4.4p of the oblate minimum are distributed
within the $f_{5/2}, p_{3/2}$ orbitals, the hole orbitals are the same as in the prolate case.
It is interesting to
observe that for small $q$ values,  the particle occupation numbers change much earlier than expected
from Fig.~\ref{Fig:SPE}. For both the proton and neutron systems already at $|3|$ barn there are
significant changes. This feature is related  with pairing and with the fact  that in Fig.~\ref{Fig:SPE} one
can not realize the \(nlj\) mixing since the lines correspond to s.p.e. of a given $K$. 
In the upper part of Fig.~\ref{Fig:occ} the proton and neutron pairing energies are plotted. As expected
the proton pairing energy is zero at and around  the spherical shape but rather large for $|q| \ge 3 $b. 
The neutron
pairing energy is very large for small $|q|$ values and decreases quite fast for large  $|q|$ values. 
We mentioned above that the oblate and the prolate minima are the two first excited states above the ground
state, the reason is that the lowest two quasiparticle states build on the ground state lie much higher, 
for protons  ($E_{p-h}^{min} = 5.7 $ MeV) due to the large s.p.e gap of the shell closure  and for neutrons 
($E_{2qp}^{min} = 3.6 $ MeV) because of the huge  pairing energy. The behavior of the pairing energies for the other 
lead isotopes are not very different from the plotted ones.

\begin{figure}
{\centering {\includegraphics[angle=0,width=\columnwidth]{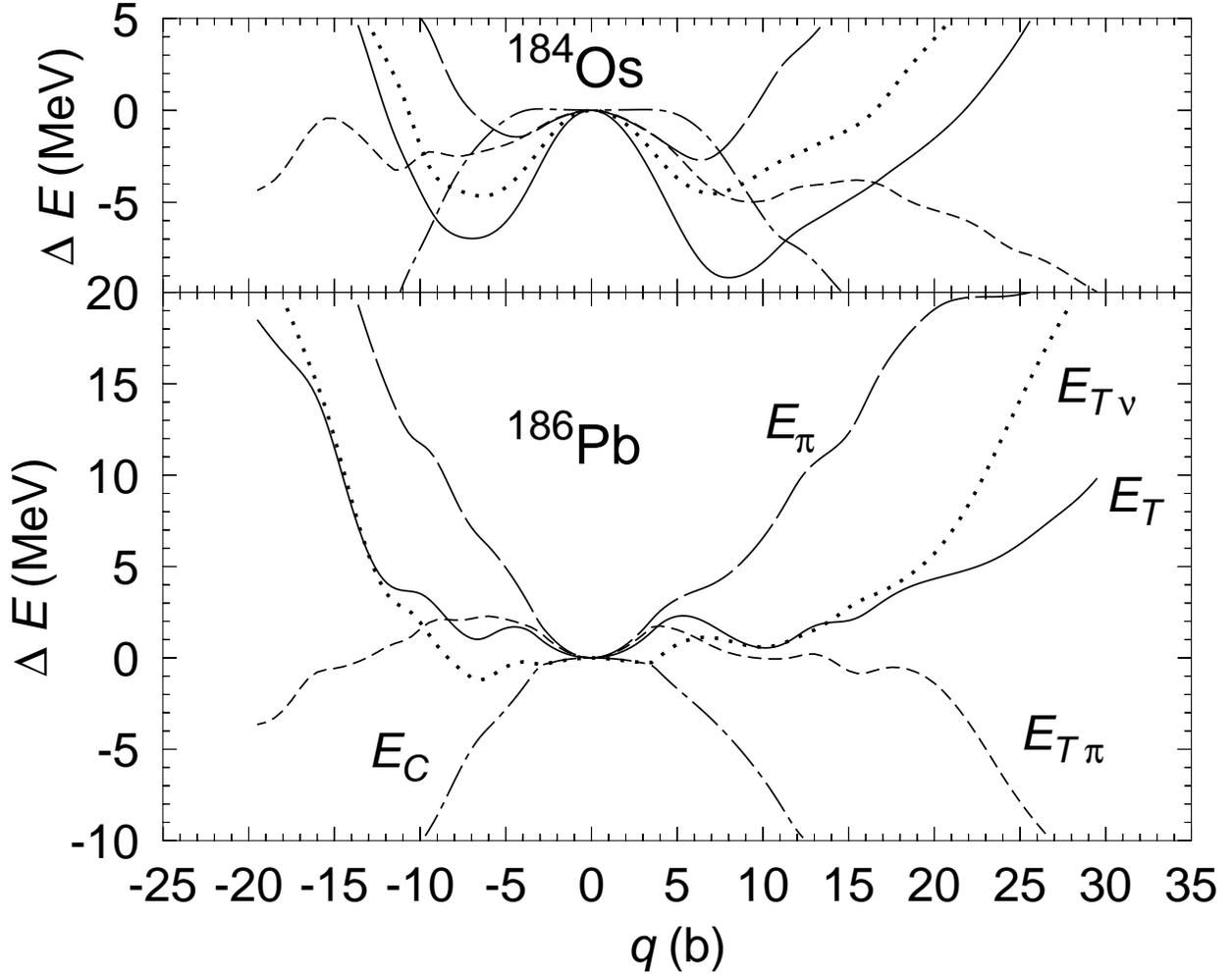}} \par}
\caption{The relative total energy $\Delta E = E(q) - E(q=0)$  and its 
different contributions as a function of $q$
for the nuclei \protect\(^{184}\)Os and \protect\(^{186}\)Pb.}
\label{Fig:Epnt186}
\end{figure}

In our study  of occupancies we got different results than the ones obtained in earlier calculations
\cite{HJM.87,NAZ.93}. These occupancies, however, do not tell us directly the mechanism to generate  
the energy minima. To unveil the origin of the
different minima let us now analyze the different energy contributions. The total HFB energy can be separated as 
$E_T = E_{T \nu} + E_{T \pi}$ for
protons and neutrons and additionally $ E_{T \pi} =  E_{\pi} + E_{C}$, with $E_{C}$ the Coulomb energy and
$E_{\pi}$ the pure nuclear energy. These energies, relative to the spherical shape,  have been plotted 
in Fig.~\ref{Fig:Epnt186}  for the nuclei \(^{184}\)Os and \(^{186}\)Pb.
 We have included the \(^{184}\)Os $(N=108, Z=76)$ results in order to illustrate the behavior of a system 
close to \(^{186}\)Pb but {\em without} the impact of the $Z=82$ shell closure.
 Let us first discuss the \(^{184}\)Os case.  
  For $E_{T \nu}$ the proximity to the mid-shell $N=104$ favors 
 deformed neutron minima as we can see in the 5 MeV deep prolate and oblate minima. The 6-hole proton 
system would prefer a spherical or a not very pronounced deformed minimum. However, due to the p-n 
interaction the proton system is driven to deformed minima about 2 to 3 MeV deep. The behavior of 
$E_{C}$ can be classically  understood, it  should be a maximum at the spherical 
shape and  rather flat around this shape corresponding to the fact that for small
deformation  parameters $\beta$ the correction to the spherical $E_{C}$ is proportional to $\beta^2$.
For quadrupole moments $ -5 b \le q \le 6.5 b $, both $E_{\pi}$ and $E_{C}$ are decreasing functions. 
The coherent addition of $E_{T \nu}$ and  $E_{T \pi}$ in this $q$ range gives
a  $E_T$ with two very well defined deformed minima. The spherical shape however remains an energy maximum.
Concerning the nucleus \(^{186}\)Pb, the energy minimum of $E_{\pi}$ is, as expected,
at the spherical shape because of the magic number $Z=82$. Small deviations around this shape increase 
this energy in a parabolic way until the single particle occupancies change in a sizable amount, see 
Fig.~\ref{Fig:occ}. At larger $q$ values one can distinguish the additional level crossings.
  $E_{T \pi}$ is rather flat for not too large deformations being 
larger than zero only for $-13  b \le q \le 13  $b.  It displays a very
shallow minimum around $q \approx  10  $b and a second one around $q \approx  16 $b.
On the oblate side we find a flat maximum (!) around $-7 $b and some level crossings at higher $q$ values.
Concerning $E_{T \nu}$  we find a very flat maximum around the spherical shape and two minima, 
one on the prolate ($q \approx  10  $b) and one on the oblate side  ($q \approx  -7  $b) as  
well as two shoulders at larger deformations. Around the spherical maximum we find a parabolic behavior for 
the same reasons as in the proton case. It is interesting to observe the presence of shoulders 
in $E_{\pi}$ at the same $q$ values where the neutron system has minima or shoulders, this is due to 
the p-n interaction.  $E_{T }$ displays three minima,
corresponding to spherical, prolate and oblate shapes. The deformed minima appear at  
exactly the same quadrupole moments as for the neutron energy. That means according to this picture
the protons  drive the system to the spherical ground state while the neutrons are  
responsible for the deformed minima. The coexistence phenomenon in the neutron deficient
lead isotopes can be viewed from a microscopic point of view as the result of the following features: 
As we have seen in the \(^{184}\)Os plot, in the absence of the shell closure the {\em proton system} will 
display a maximum at the spherical shape and two deformed minima and so will do the {\em neutron system}.
In the presence of a strong  proton shell closure, $E_{\pi}$ together with $E_{C}$
provide a spherical minimum and a rather shallow  $E_{T \pi}$ surface at $-15 {\rm b} \le q  \le 20  $b. 
Furthermore, the p-n interaction drives the neutron system to 
a spherical  shape providing  a very flat $E_{T \nu}$ surface around zero quadrupole moment though  not
a minimum.  This energy flattening causes that the prolate-spherical and spherical-oblate energy barriers become 
very small. Notice also that  due to the strong shell closure  $E_{\pi}$ grows faster that $E_{C}$  decreases 
around the spherical shape (contrary to \(^{184}\)Os).  This effect causes  in \(^{186}\)Pb a  
cancellation  of  $E_{T \pi}$ with $E_{T \nu}$ in the prolate and oblate minima in such a 
way that the deformed minima  of  $E_{T}$ lie close to the spherical minimum. From the previous discussion
we can infer that the existence of spherical, prolate and oblate states within 1 MeV  is  a very special 
characteristic  of  this mass region favored by 
the mid-shell $N=104$ neutron configuration and the proton shell closure. The p-n
and the pairing interactions provide the remaining ingredients.

\begin{figure}
{\centering {\includegraphics[angle=0,width=\columnwidth]{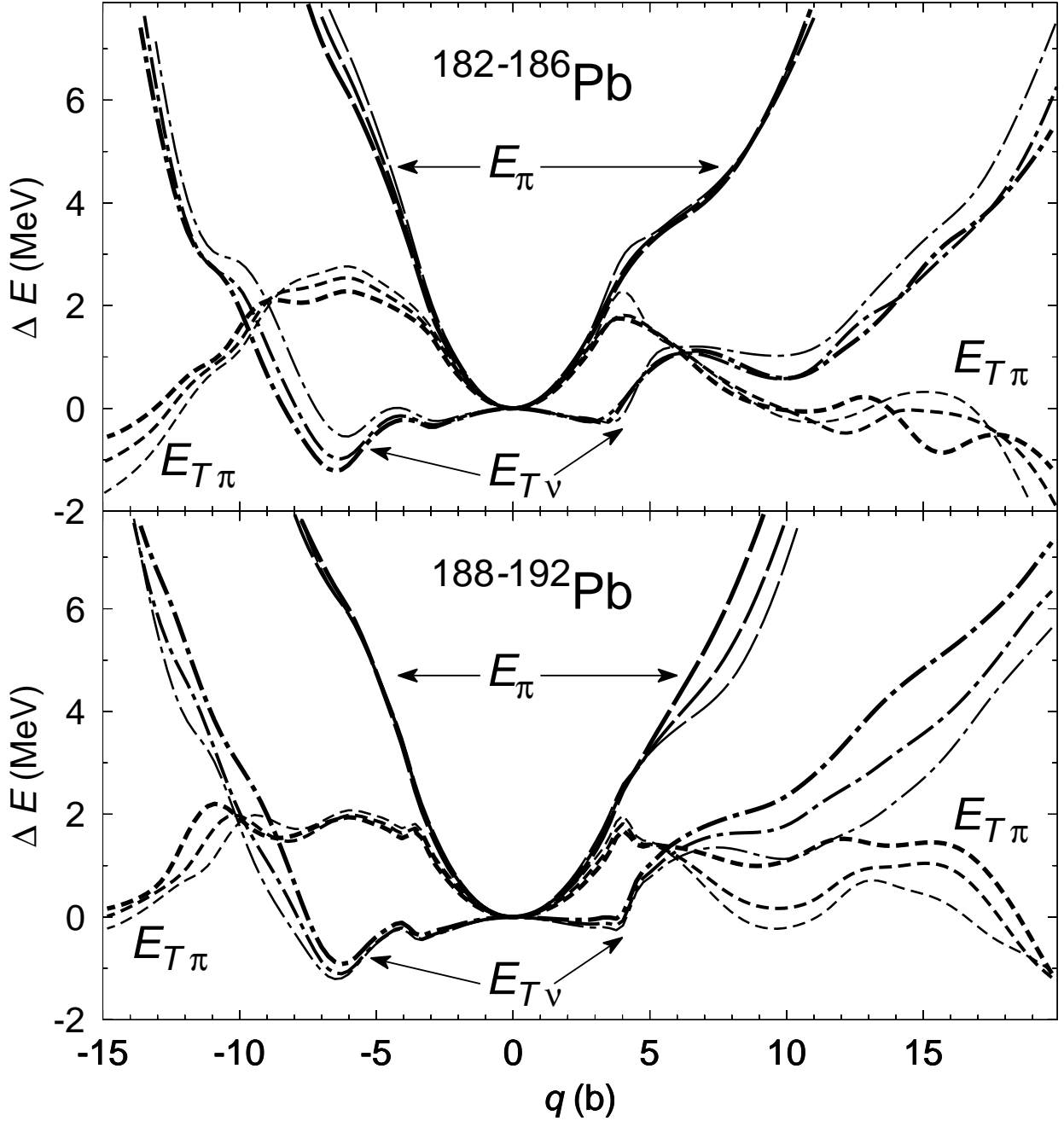}} \par}
\caption{Same as Fig.\ref{Fig:Epnt186} but for \protect\(^{182-192}\)Pb.
In both panels the line thickness increase with the neutron number.}
\label{Fig:MFPESNP}
\end{figure}

We shall now try to understand the evolution of the shape coexistence with the neutron number.
In Fig.~\ref{Fig:MFPESNP} the different energies for the nuclei $^{182-192}$Pb are plotted.
The $E_{C}$'s are not drawn because for all isotopes they look rather similar to the one depicted
in Fig.~\ref{Fig:Epnt186} for $^{186}$Pb.
In Fig.~\ref{Fig:MFPESNP} we can follow very clearly the evolution of the prolate and oblate minima. In the
upper part for the nuclei with a prolate first excited state, i.e., \(^{182-186}\)Pb, we can
observe how  the neutron oblate well gets deeper with increasing neutron number.  The prolate
minimum does the same though to a lesser extend. This can be understood looking at 
Fig.~\ref{Fig:SPE}, where we can see on the oblate side, around $q= -7 $ b, that 
for  neutron numbers $N=100,102$ and $104$ we are filling down-sloping $i_{13/2}$ orbitals. On the
prolate side and around $q = 10 \;$b, there are down- and up-sloping orbitals providing a smaller change in the
energy.  In the lower part of Fig.~\ref{Fig:MFPESNP} we have plotted the
energies for those nuclei with an oblate first excited state, i.e., \(^{188-192}\)Pb. With increasing
neutron number we
observe a small decrease of the oblate well and the disappearance of the prolate well. This
behavior can be understood again looking at Fig.~\ref{Fig:SPE}, there we find that for
neutron number $N=106,108$ and $110$ and around $q = 10 \;$b we are filling different 
up-sloping  orbitals. On the oblate side, on the other hand,  we are filling on the average 
orbitals without a well defined behavior. Since the behavior around the spherical minimum,
$-5 b \le q \le 5 $b, is dominated by the proton behavior through the p-n
interaction, $E_{T \nu}$  does not change in this region.
Concerning the $E_{\pi}$ we find that
as expected they do not vary as much as the neutron ones. We can furthermore observe the effect
of the p-n interaction~: In the places where $E_{T \nu}$ presents stronger
variations so does $E_{\pi}$, see the oblate side in \(^{182-186}\)Pb and the prolate
side in \(^{188-192}\)Pb. $E_{T \pi}$ does not show a clear behavior on the oblate side.
 Around the prolate minimum and for the nuclei 
\(^{182-186}\)Pb it shows flat minima and for \(^{188-192}\)Pb somewhat more pronounced ones. This behavior
as explained above is caused to the p-n interaction. The addition of $E_{T \pi}$ and  $E_{T \nu}$  provides 
 $E_{T}$ already discussed in Fig.\ref{Fig:MFPES}.

In conclusion  the shape coexistence phenomenon in the lead isotopes
has been analyzed in the selfconsistent HFB approach with the Gogny force. We have found a reasonable good description
of the experimental energies as well as clear explanations for the existence of the three low-lying states
of different shapes and of the evolution of these minima with the mass number. In particular, we have
shown that the ``traditional" understanding of shape coexistence in the lead isotopes 
based on proton excitations has to be modified because the  proton potentials  display either
maxima or very shallow minima at the corresponding quadrupole moments.

This work was supported in part by DGI, Ministerio de Ciencia y Tecnologia, Spain, 
under Project BFM2001-0184.


\end{document}